\documentstyle[preprint,eqsecnum,aps]{revtex}
\begin{document}
\preprint{MAD/TH/93-05}
\title{Casimir operators of the exceptional group $G_2$}
\author{A.M. Bincer and K. Riesselmann}
\address{Department of Physics, University of Wisconsin-Madison,
Madison, WI 53706}
\maketitle
\begin{abstract}
	We calculate the degree 2 and 6 Casimir operators of $G_2$ in
	explicit form, with the generators of $G_2$ written in terms of
	the subalgebra $A_2$.
\end{abstract}
\begin{flushleft}

PACS: 02.20.+b

\end{flushleft}
\newpage

\section{Introduction}

The general algorithm for obtaining invariant quantities in
semi-simple groups goes back to Killing \cite{Kill}, Cartan
\cite{Cart}, and Racah \cite{setinv}. Killing's procedure is based
on the characteristic equation defined by the adjoint
representation. Racah extended this idea using any representation,
and subsequent work showed that a complete set of invariants can be
obtained in this way \cite{compl,compl2}.

Our motivation for the work presented here are the invariants of
the exceptional group $E_6$ which are of particular interest to
particle physics. To set up the method, it is illustrative to look
at the Casimir operators of $G_2$ using generators in terms of the
subalgebra $A_2$. (The group $E_6$ has the subalgebra $A_2\oplus
A_2\oplus A_2$.)

\section{Generators of $G_2$ in $A_2$ basis}

We identify the 14 generators of $G_2$ in terms of the $A_2$
subalgebra. Corresponding to the reduction
%
%
\begin{equation}
\label{eq2.1}
{\bf 	14=8+3+\bar3}
\end{equation}
under the restriction of $G_2$ to $A_2$ we label the generators as
${g_k}^\ell,\ a_k,\ b^\ell$ with all indices ranging from 1 to 3. We
have the commutation relations \cite{commu,rep}
%
\begin{eqnarray}
\label{eq2.2}
	\left[ {g_k}^\ell,\ {g_m}^n \right] &=& {\delta_m}^\ell
{g_k}^n-{\delta _k}^n {g_m}^\ell\\
\label{eq2.3}
	\left[ {g_k}^\ell,\ a_m \right] &=& {\delta_m}^\ell
	a_k-\frac{1}{3}{\delta _k}^\ell a_m\\
	\label{eq2.4}
	\left[ {g_k}^\ell,\ b^n \right] &=& -{\delta_k}^n
	b^\ell+\frac{1}{3}{\delta _k}^\ell b^n\\
	\label{eq2.5}
	\left[ a_m,\ b^n \right] &=& {g_m}^n\\
	\label{eq2.6}
	\left[ a_m,\ a_n \right] &=& -\frac{2}{\sqrt{3}}\epsilon
	_{mn\ell}b^\ell\\
	\label{eq2.7}
	\left[ b^m,\ b^n \right] &=& \frac{2}{\sqrt{3}}\epsilon
	^{mn\ell}a_\ell
\end{eqnarray}
Equation (\ref{eq2.2}) together with the constraint
%
%
\begin{equation}
\label{eq2.8}
	\sum_{k}^{}{g_k}^k=0
\end{equation}
identifies ${g_k}^\ell$ as the octet of generators generating $A_2$.
Equation (\ref{eq2.3}) states that the three generators
$a_m$ transform as a triplet,
and Eq.\ (\ref{eq2.4}) states that the $b^n$ transform as an
antitriplet under the $A_2$ generated by the ${g_k}^\ell$. These
properties can also be read off the root diagram shown in Fig.\
\ref{fig1}.
Lastly we note that in a unitary representation the hermiticity
properties are given by
%
%
\begin{eqnarray}
\label{eq2.9}
	{g_k}^{\ell\dagger} &=& {g_\ell}^k,\\
	\label{eq2.10}
	{a_m}^\dagger  &=& b^m.
\end{eqnarray}

\section{The Matrix Q}

Consider an arbitrary finite-dimensional semisimple Lie algebra
$L$ and denote its generators by $X_\mu,\ \mu=1,2,\ldots,d$,
where $d={\rm dim} L$. Consider next some nontrivial representation
and denote the generators in that represention by $x_\mu$. Because
$L$ is semisimple it follows that the $d\times d$ symmetric
matrix $g_{\mu \nu }$ defined by
%
%
\begin{equation}
\label{eq3.1}
	g_{\mu \nu }=tr(x_\mu x_\nu )
\end{equation}
is nonsingular. Hence we can define
%
%
\begin{equation}
\label{eq3.2}
	g^{\mu \nu }=\left( g^{-1} \right)_{\mu \nu }
\end{equation}
and introduce the matrix $Q$ by
%
%
\begin{equation}
\label{eq3.3}
	Q=X_\mu \otimes x^\mu =g^{\mu \nu }X_\mu \otimes x_\nu .
\end{equation}

We also can define the $(p+1)$-th power of $Q\ (p\geq0)$ by
%
%
\begin{equation}
\label{eq3.4}
	Q^{p+1}=QQ^p
\end{equation}
with the property
%
%
\begin{equation}
\label{eq3.5}
	\left[ X_\mu ,\ tr\,Q^p \right]=0
\end{equation}
so that degree $p$ Casimir operators can be taken equal to
$tr\,Q^p$. The procedure just described is essentially that of
Gruber and O'Raifeartaigh \cite{compl} where the
proof of Eq.\ (\ref{eq3.5}) can be found.

We now construct $Q$ in the case of $L=G_2$. The nontrivial
representation is chosen as the 7-dimensional irreducible
representation of $G_2$ (the smallest irreducible representation of
$G_2$). Thus $Q$ is a $7\times 7$ matrix whose matrix elements are
proportional to the 14 generators of $G_2$. Since the ${\bf 7}$
of $G_2$ decomposes under $A_2$ as
%
%
\begin{equation}
\label{eq3.6}
	{\bf 7}={\bf 3}\oplus{\bf 1}\oplus{\bf \bar3}
\end{equation}
the $7\times 7$ matrix $Q$ naturally breaks up into a $3\times 3$
matrix given as
%
%
\begin{equation}
\label{eq3.7}
	Q=\left(
	\begin{array}{ccc}
		G &\alpha &B\\
		\beta ^T &0 &-\alpha ^T\\
		A &-\beta  &-G^T
	\end{array}
	 \right).
\end{equation}
Here $A,\ B,\ G,\ G^T$ are $3\times 3$ matrices, $\alpha ,\
\beta $ are $3\times 1$ (column) matrices, and $\alpha ^T,\ \beta ^T$
are $1\times 3$ (row) matrices. Explicitly, we have
%
%
\begin{equation}
\label{eq3.8}
	\left.
	\begin{array}{r}
		G_{k\ell}={g_k}^\ell,\quad (G^T)_{k\ell}={g_\ell}^k\\
		A_{k\ell}=-\frac{1}{\sqrt{3}}
		\epsilon ^{k\ell r}a_r,\ B_{k\ell}=\frac{1}{\sqrt{3}}\epsilon
		_{k\ell r}b^r\\
		\alpha _k=\sqrt{\frac{2}{3}}a_k,\quad
		\beta _k=\sqrt{\frac{2}{3}}b^k
    \end{array}
	\right\}
\end{equation}

Except for some renumbering of rows and columns the
representation of the generators of $G_2$ by $7\times 7$ matrices
used in Eq.\ (\ref{eq3.7}) is precisely the same as given by
Patera \cite{Patera} and
it agrees with
Berdjis \cite{compl2}
and Ekins and Cornwell \cite{sub} but for different normalization
of the generators.

We write the $7\times 7$ matrix $Q^p$ as
%
%
\begin{equation}
\label{eq3.9}
	Q^p=\left(
	\begin{array}{ccc}
		G_p &\alpha _p &B_p\\
		\beta _p^T &c_p &\gamma _p^T\\
		A_p &\rho _p &H_p
	\end{array}
	 \right)
\end{equation}
where $A_p,\ B_p,\ G_p\ H_p$ are $3\times 3$ matrices, $\alpha
_p,\ \beta _p,\ \gamma _p,\ \rho _p$ are $3\times 1$ matrices, and
$c_p$ is a $1\times 1$ matrix. Eq.\ (\ref{eq3.4}) implies
%
%
\begin{eqnarray}
\label{eq3.10}
	G_{p+1} &=& GG_p+\alpha \beta _p^T+BA_p,\nonumber \\
	\alpha _{p+1} &=& G\alpha _p+\alpha c_p+B\rho _p,\nonumber \\
	B_{p+1} &=& GB_p+\alpha \gamma _p^T+BH_p,\nonumber \\
	\beta _{p+1}^T &=& \beta ^TG_p-\alpha ^TA_p,\nonumber \\
	c_{p+1} &=& \beta ^T\alpha _p-\alpha ^T\rho _p,\nonumber \\
	\gamma _{p+1}^T &=& \beta ^TB_p-\alpha ^TH_p,\nonumber \\
	A_{p+1} &=& AG_p-\beta \beta _p^T-G^TA_p,\nonumber \\
	\rho _{p+1} &=& A\alpha _p-\beta c_p-G^T\rho _p,\nonumber \\
	H_{p+1} &=& AB_p-\beta \gamma _p^T-G^TH_p.
\end{eqnarray}
These recursive definitions enable us to investigate the explicit
form of the Casimir operators of $G_2$ in the next section.

\section{The Casimir operators ${\cal C}_{\lowercase{p}}$}

It follows from the preceding that the degree $p$ Casimir operator
${\cal C}_p$ is given by
%
%
\begin{equation}
\label{eq4.1}
	{\cal C}_p\equiv tr Q^p=c_p+tr(G_p+H_p)
\end{equation}
which by repeated use of Eq.\ (\ref{eq3.10}) can be explicitly
exhibited as a homogeneous polynomial of degree $p$ in the matrix
elements of $Q$, i.e., the generators of $G_2$.

Functionally independent Casimir operators are in fact only two:
${\cal C}_2$ and ${\cal C}_6$. This is because the number of
independent Casimir operators equals the rank of the group and their
degrees obey the relation
$p=m+1$, where $m$ are the so-called exponents of the group.
For $G_2$ we have
that rank = 2 and $m=1$ and 5.

In our formulation we find the functionally independent Casimir
operators as follows: Since $Q$ is a $7\times 7$ matrix, all
Casimirs of degree $p>7$ are, by the Cayley-Hamilton theorem, not
independent of Casimirs of lower degree. Further, $Q$ carries an
antisymmetry property corresponding to the fact that all irreducible
representations of $G_2$ are orthogonal---as a consequence all odd
degree Casimirs are expressible in terms of lower degree (and
ultimately even degree) Casimirs. Thus we need only evaluate ${\cal
C}_2,\ {\cal C }_4$ and ${\cal C}_6$, and we will find that ${\cal
C}_4$ is expressible in terms of ${\cal C}_2$.

For $p=2$  Eq.\ (\ref{eq4.1}) becomes
%
%
\begin{eqnarray}
\label{eq4.2}
	{\cal C}_2 &=& c_2+tr(G_2+H_2)\nonumber \\
	 &=& \beta ^T\alpha +\alpha ^T\beta +tr(GG+\alpha \beta
	 ^T+BA+AB+\beta \alpha ^T+G^TG^T)\nonumber \\
	  &=& 2\left\{ \langle g^2\rangle
	  +a\cdot b+b\cdot a \right\}\nonumber \\
	   &=& 2\left\{ \langle g^2\rangle +2b\cdot a \right\}
\end{eqnarray}
where we used the following identities and definitions:
%
%
\begin{eqnarray}
\label{eq4.3}
	\alpha ^T\beta   &=& \beta
	^T\alpha =
	\frac{2}{3}b^ka_k\equiv\frac{2}{3}b\cdot a,\nonumber \\
	tr G^T G^T  &=& tr GG = {g_k}^\ell {g_\ell}
	^k\equiv \langle g^2\rangle,\nonumber \\
	tr \beta \alpha ^T  &=&  tr \alpha \beta ^T =
	\frac{2}{3}a_kb^k\equiv \frac{2}{3}a\cdot b\nonumber \\
	tr AB  &=&
	trBA=B_{k\ell}A_{\ell k}=\frac{1}{3}\epsilon _{k\ell
	r}b^r\epsilon ^{k\ell s}a_s=\frac{2}{3}b\cdot a.
\end{eqnarray}

We remark that $\langle g^2\rangle$ and $a\cdot b=b\cdot a$ are the {\em
only\/} quadratic structures that can be formed out of the octet
${g_k}^\ell$, the triplet $a_k$ and the antitriplet $b^k$ that are
SU(3) invariants---hence the quadratic Casimir of $G_2$ must be
some linear combination of $\langle g^2\rangle$ and $b\cdot a$.
Our second remark is that, since $a\cdot b=b\cdot a$, the
quadratic Casimir is a {\em symmetric\/}
polynomial in the generators, homogeneous of degree 2.

For $p=4$ we have from Eq.\ (\ref{eq4.1})
%
%
\begin{eqnarray}
\label{eq4.4}
	{\cal C}_4  &=& c_4+tr(G_4+H_4)\nonumber \\
	 &=& tr(G^4+\mbox{transpose})\nonumber \\
	 &&+\ tr\left( \frac{1}{2}ABAB+\alpha \beta ^T\alpha \beta
	 ^T+\mbox{transpose} \right)_{\mbox{cycle 2}}\nonumber \\
	 &&+\ tr\left( G^2\alpha \beta ^T+G^2BA-\frac{1}{2}GBG^TA+GB^T\beta
	 \beta ^T\right.\nonumber \\
	 &&\phantom{+\ tr+}
	 +\left.A^TG\alpha \alpha ^T+BA\alpha \beta
	 ^T+\frac{1}{2}\alpha \alpha ^T\beta \beta ^T+\mbox{transpose}
	 \right)_{\mbox{cycle 4}}
\end{eqnarray}
where the ``transpose" instruction means: add the transpose of all
the preceding terms in the bracket. The ``cycle" instruction
means: add all cyclic permutations. (Factors $\frac{1}{2}$ were
inserted to take care of those cyclic
terms that are equal to their own
transpose.) Thus e.g.
%
%
\begin{eqnarray}
\label{eq4.5}
	G^4+\mbox{transpose} &\equiv& G ^4+G ^{T4},\\
	\label{eq4.6a}
	tr(\alpha \beta ^T\alpha \beta ^T)_{\mbox{cycle 2}} &\equiv
	& (\alpha
	_a\beta _b\alpha _b\beta _a)_{\mbox{cycle 2}}\nonumber \\
	 &\equiv
	 & \alpha _a\beta _b\alpha _b\beta _a+\beta _b\alpha
	 _b\beta _a\alpha _a,\\
\label{eq4.6}
	tr(BA\alpha \beta ^T)_{\mbox{cycle 4}}&\equiv&\left(
	B_{ab}A_{bc}\alpha _c\beta _a \right)_{\mbox{cycle 4}}\nonumber\\
	 &\equiv&
	 B_{ab}A_{bc}\alpha _c\beta  _a+A_{bc}\alpha  _c\beta
	 _aB_{ab}+\alpha _c\beta _aB_{ab}A_{bc}+\beta
	 _aB_{ab}A_{bc}\alpha _c.
\end{eqnarray}

Eq.\ (\ref{eq4.4}) exhibits ${\cal C}_4$ as a homogeneous polynomial
of degree 4 in the generators of $G_2$. Alternatively, using
commutation relations we can combine terms containing the same
generators in different orders, and we arrive at an expression for
${\cal C}_4$ that shows that it is a function of ${\cal C}_2$:
%
%
\begin{equation}
\label{eq4.7}
	{\cal C}_4={\cal C}_2\left[
	\frac{1}{4}{\cal C}_2+\frac{14}{3} \right].
\end{equation}
This result includes the fact that
%
%
\begin{equation}
\label{eq4.8.a}
	G^3=3G^2+\left( \frac{1}{3}trG^3-trG^2 \right){\bf 1}+\left(
	\frac{1}{2}trG^2-2 \right)G,
\end{equation}
which follows from the generalized Cayley-Hamilton theorem.

We finally consider $p=6$. This Casimir operator will be given in
two different forms. In the first form we exhibit it explicitly as
a homogeneous polynomial of degree 6 in the generators possessing
cyclic symmetry, by which we mean that if a product of 6
generators appears in a certain order then so do all the cyclic
permutations of that order. We exploit this cyclic property in the
notation (as was already seen for $p=4$) to write out fairly
economically the 416 terms that arise.

In the second form we use commutation relations to relate terms
that only differ in the order in which the generators appear. Here
we end up with fewer terms but at the cost of having a polynomial
of degree six which is no longer homogeneous.

To keep track of the 416 terms in the first form we group them by
the degree in $G$ and $G^T$. Thus, ${\cal C}_6$ is a sum of: \\
the 2
terms of degree 6 in $G$ and $G^T$:
%
%
\begin{equation}
\label{eq4.8}
	tr(G^6+\mbox{transpose})
\end{equation}
{\em plus\/} the $7\cdot 6$ terms of degree 4 in $G$ and $G^T$:
%
%
\begin{equation}
\label{eq4.9}
	tr\left( G^4\alpha \beta ^T+
	G^4BA+G^3BG^TA^T+\frac{1}{2}G^2BG^{T2}A+\mbox{transpose}
	\right)_{\mbox{cycle 6}}
\end{equation}
{\em plus\/} the $8\cdot6$ terms of degree 3 in $G$ and $G^T$:
%
%
\begin{equation}
\label{eq4.10}
	tr(G^3\alpha \alpha ^TA^T+G^2\alpha \alpha ^TG^TA+G^3B^T\beta
	\beta ^T+G^2BG^T\beta \beta ^T+\mbox{transpose})_{\mbox{cycle 6}}
\end{equation}
{\em plus\/} the $21\cdot6+4\cdot3=138$ terms of degree 2 in $G$ and
$ G^T$:
%
%
\begin{eqnarray}
\label{eq4.11}
	tr&&(G^2\alpha \beta ^T\alpha \beta ^T+G^2BA\alpha \beta
	^T+G^2\alpha \beta ^TBA+G^2BABA\nonumber \\
	&&+G^2\alpha \alpha ^T\beta \beta ^T+G^2B\beta \alpha
	^TA+GBAG\alpha \beta ^T+GBG^TA^T\alpha \beta ^T\nonumber \\
	&&+G\alpha \beta
	^TBG^TA^T+\frac{1}{2}GBA^TBG^TA+\frac{1}{2}GBG^TABA^T\nonumber \\
	&&+\frac{1}{2}G\alpha \alpha ^TG ^T\beta \beta
	^T+\mbox{transpose})_{\mbox{cycle 6}}\nonumber \\
	+tr&&(G\alpha \beta ^TG\alpha \beta
	^T+GBAGBA+\mbox{transpose})_{\mbox{cycle 3}}
\end{eqnarray}
{\em plus\/} the $20\cdot6$ terms of degree 1 in $G$ and $G^T$:
%
%
\begin{eqnarray}
\label{eq4.12}
	tr&&(G\alpha \beta ^T\alpha \alpha ^TA^T+G\alpha \alpha ^T\beta
	\alpha ^TA^T+G\alpha \alpha ^TA^T\alpha \beta ^T\nonumber \\
	&&+GB^T\beta \beta ^T\alpha \beta ^T+GB ^T\beta \alpha
	^T\beta \beta ^T+G\alpha \beta ^TB^T\beta \beta ^T\nonumber \\
	&&+
	GB^TA\alpha \alpha ^TA+G\alpha
	\alpha ^TAB^TA+GB\beta \beta ^TBA^T\nonumber \\
	&&+GBA^TB\beta \beta ^T+\mbox{transpose})_{\mbox{cycle 6}}
\end{eqnarray}
{\em plus\/} lastly the $3\cdot2+2\cdot3+9\cdot6=66$ terms of degree
0 in $G$ and $G^T$:
%
%
\begin{eqnarray}
\label{eq4.13}
	tr&&(\frac{1}{2}ABABAB+\alpha \beta ^T\alpha \beta ^T\alpha \beta
	^T+\mbox{transpose})_{\mbox{cycle 2}}\nonumber \\
	+tr&&(\beta \beta ^TB\beta \beta ^TB+\alpha \alpha ^TA\alpha
	\alpha ^TA )_{\mbox{cycle 3}}\nonumber \\
	+tr&&(\alpha \beta ^T\alpha \beta ^TBA+\alpha \alpha ^T\beta
	\beta ^TBA+\alpha \beta ^TBABA\nonumber \\
	&&+\alpha \alpha ^T\beta \alpha ^T\beta \beta
	^T+\frac{1}{2}\alpha \beta ^TB\beta \alpha
	^TA+\mbox{transpose})_{\mbox{cycle 6}}
\end{eqnarray}

On the other hand, using the definitions given in Eq.\ (\ref{eq4.3})
and applying commutation relations, we can rewrite
${\cal C}_6$ as follows \cite{Reduce}:
%
%
\begin{eqnarray}
\label{eq4.14}
	{\cal C}_6 &=& \frac{44}{9}a\!\cdot\! b\,a\!\cdot\! b\,a\!\cdot\!
b+\frac{10}{3}a\!\cdot\! b\,a\!\cdot\! b\langle
g^2\rangle-2b\!\cdot\!
g\!\cdot\! a\,b\!\cdot\! g\!\cdot\! a\nonumber \\
	&&+8b\!\cdot\! g\!\cdot\! g\!\cdot\! a\,a\!\cdot\! b+5a\!\cdot\!
b\langle g^2\rangle\langle g^2 \rangle-4b\!\cdot\! g\!\cdot\!
g\!\cdot\! a\langle g^2\rangle\nonumber \\
	&&+4b\!\cdot\! g\!\cdot\! a\langle g^3\rangle+\frac{2}{3}\langle
g^3\rangle\langle g^3\rangle+\frac{1}{2}\langle g^2\rangle\langle
g^2\rangle\langle g^2\rangle\nonumber \\
	&&-\frac{8}{\sqrt{3}}\epsilon ^{ijk}a_i(g\!\cdot\! a)_j(g\!\cdot\!
g\!\cdot\! a)_k-\frac{8}{\sqrt{3}}\epsilon _{ijk}b^i(b\!\cdot\!
g)^j(b\!\cdot\! g\!\cdot\! g)^k\nonumber \\
	&&-20a\!\cdot\! b\,b\!\cdot\! g\!\cdot\! a-\frac{20}{3}\langle
g^3\rangle a\!\cdot\! b+\frac{22}{\sqrt{3}}\langle g^2\rangle
b\!\cdot\! g\!\cdot\! a-2\langle g^2\rangle\langle
g^3\rangle\nonumber \\
 &&-\frac{112}{9}b\!\cdot\! g\!\cdot\! g\!\cdot\! a+88a\!\cdot\!
b\langle g^2\rangle+\frac{836}{9}a\!\cdot\! b\,a\!\cdot\!
b+21\langle g^2\rangle\langle g^2\rangle\nonumber \\
	&&+\frac{5208}{81}b\!\cdot\! g\!\cdot\! a-\frac{40}{3}\langle
g^3\rangle\nonumber \\
	&&+\frac{584}{81}a\!\cdot\! b+\frac{3188}{27}\langle g^2\rangle.
\end{eqnarray}

Depending on the application one might prefer the version given by
Eqs.\ (\ref{eq4.8}-\ref{eq4.13}); although there are 416 terms
they are all of the same degree and exhibit cyclic symmetry. The
version given by Eq.\ (\ref{eq4.14}) on the other hand has only 23
terms but is no longer homogeneous of degree 6 in the generators
and no longer displays its explicit symmetry in the generators.

Previously, the only other fully explicit expression for ${\cal
C}_6$ in the literature has been given by Hughes and Van der Jeught
\cite{AnsG2}. Their generators of $G_2$ are given in the $A_1\oplus
A_1$ basis (in contrast to our use of the $A_2$ basis) and they
obtain for ${\cal C}_6$ an expression involving 29 terms. The
$A_1\oplus A_1$ basis, however, has the disadvantage of not
displaying as much symmetry in the generators. Our
motivation is to find explicit expressions for the Casimirs of
$E_6$, and
we expect the subalgebra $A_2\oplus A_2\oplus A_2$ to give the
simplest results. The studies of $G_2$ in terms of $A_2$ seem to
foster these hopes.

Lastly we should like to mention the work of Englefield and King
\cite{King}, in which explicit expressions are given for the {\em
eigenvalues\/} of all the Casimir operators of $G_2$.

\acknowledgements

One of the authors (AMB) wishes to express his thanks to Charlie
Goebel for numerous discussions.  The other author (KR)
acknowledges the support of DOE grant No.\ DE-AC02-76-ER00881.

\begin{figure}
	\caption{The root diagram of $G_2$ in $A_2$ basis.}
	\label{fig1}
\end{figure}


\begin{references}
\bibitem{Reduce} To carry out this step of our calculation we
	 used the algebraic programming system REDUCE.
\end{references}
\end{document}